\documentclass[twocolumn,showpacs,prl]{revtex4}

\usepackage{graphicx}

\newcommand{\figurewidth}{0.42\textwidth}

\begin{document}

\title{Salt-induced collapse and reexpansion of highly charged flexible
polyelectrolytes}
\author{Pai-Yi Hsiao}
\author{Erik Luijten}
\email[Corresponding author. E-mail: ]{luijten@uiuc.edu}
\affiliation{%
    Department of Materials Science and Engineering and
    Frederick Seitz Materials Research Laboratory,\\
    University of Illinois at Urbana-Champaign,
    Urbana, Illinois 61801, U.S.A.
}

\date{May 22, 2006}

\begin{abstract}
  We study the salt-dependent conformations of dilute flexible
  polyelectrolytes in solution via computer simulations. Low
  concentrations of multivalent salt induce the known conformational
  collapse of individual polyelectrolyte chains, but as the salt
  concentration is increased further this is followed by a
  \emph{reexpansion}.  We explicitly demonstrate that multivalent
  counterions can overcompensate the bare charge of the chain in the
  reexpansion regime.  Both the degree of reexpansion and the occurrence
  of overcharging sensitively depend on ion size. Our findings are
  relevant for a wide range of salt-induced complexation phenomena.
\end{abstract}


\maketitle

Many biological and synthetic polyelectrolytes undergo two macroscopic
phase transitions upon addition of multivalent salt or charged small
molecules~\cite{olvera95,pelta96a,raspaud98,sabbagh99,sanders05}.
First, precipitation of polyelectrolytes occurs if the concentration of
the added salt exceeds a critical value.  This precipitate subsequently
redissolves if the concentration is increased beyond a second critical
value.  These phenomena, jointly referred to as \emph{reentrant
condensation}~\cite{nguyen00}, have attracted considerable attention
because they are a fundamental and generic aspect of polyelectrolyte
behavior, with potential relevance for the understanding and development
of biological phenomena and applications such as gene
delivery~\cite{vijayanathan02}.  Under dilute conditions, the formation
and redissolution of multimolecular aggregates are replaced by,
respectively, single-chain ``collapse'' and reexpansion.  This behavior
is directly observed in, e.g., the stretching of individual DNA coils in
salty solution~\cite{murayama03}.

Whereas several theoretical explanations for reentrant condensation have
been proposed~\cite{nguyen00,solis00,solis01,solis02}, important open
questions remain regarding the suggested mechanisms, and several
predictions have only partially been verified.  Two main scenarios can
be distinguished.  (1)~\emph{Charge inversion:} Counterions form a
strongly correlated liquid at the polyelectrolyte surface and, at high
salt concentration, overcompensate the bare chain charge. Condensation
of a polyelectrolyte is predicted to take place if its effective charge
is close to zero~\cite{nguyen00,grosberg02}.  (2)~\emph{The two-state
model}~\cite{solis00} explains precipitation and redissolution of
flexible polyelectrolytes by assuming that individual chains adopt
either a collapsed or an extended state.  This scenario predicts a
strong sensitivity to ion size, which is not considered in
Refs.~\cite{nguyen00,grosberg02}. In addition, in this case overcharging
does not occur by necessity, but may be mitigated by coion
association~\cite{solis02}.

Computer simulations offer the possibility to study this behavior at a
microscopic level, but hitherto most simulations of flexible chains have
focused on solutions with only divalent salt~\cite{stevens98,liu03} or
no salt at all~\cite{stevens95,winkler98,stevens01,limbach03}.  Whereas
chain collapse has been observed under the influence of divalent salt or
counterions~\cite{winkler98}, the reexpansion observed in experiments
and predicted by theory has not been reproduced in computer
simulations.  In this Letter, we aim to clarify this apparent
discrepancy and to understand the behavior of strong polyelectrolytes in
solutions of multivalent salt by studying the conformations of dilute
chains over a wide range of salt concentrations. Our calculations allow
us to explicitly address the theoretical predictions.  Specifically, we
answer the following questions: (i)~How does the conformation of a
polyelectrolyte depend on salt concentration and valency, and under what
conditions can an extended (i.e., noncollapsed) structure be observed at
high concentrations?  (ii)~Is there a dependence on ion size, as
suggested by some theories?  (iii)~Can the charge of a polyelectrolyte
be overcompensated by condensed ions, such that its effective charge is
\emph{reversed} compared to the bare charge?

Following Stevens and Kremer~\cite{stevens95}, we model an anionic
polyelectrolyte as a bead--spring model in a continuous medium of
uniform dielectric constant~$\varepsilon$, representing the solvent.  A
chain consists of $N$ monomers of size $\sigma$, which each carry a
charge~$-e$. Adjacent monomers are connected by a FENE bond
potential~\cite{stevens95} with maximum extension $R_0=2\sigma$ and
spring constant $k=7\varepsilon_{\rm LJ}/\sigma^2$.  The counterions are
monovalent, whereas the added salt dissociates into $Z$-valent cations
and monovalent anions.  Electrostatic interactions are computed via the
Ewald summation and the excluded volume of monomers and ions is modeled
via the shifted-truncated Lennard-Jones potential with coupling constant
$\varepsilon_{\rm LJ} = k_{\rm B}T/1.2$, where $k_{\rm B}$ denotes
Boltzmann's constant and $T$ the absolute temperature. Periodic
boundaries are imposed.  We express the Coulomb interaction between
particles $i$ and~$j$ as $k_{\rm B}T\lambda_{\rm B} z_i z_j/r$, where
$\lambda_{\rm B} \equiv e^2/(4\pi\varepsilon\varepsilon_0 k_{\rm B}T)$
is the Bjerrum length, $\varepsilon_0$ the vacuum permittivity, and
$z_i$ and $z_j$ the valencies of the particles. In aqueous solution at
room temperature, $\lambda_{\rm B}=7.14$~\AA. We set $\lambda_{\rm B} =
3\sigma$, representative of a prototypical flexible polyelectrolyte such
as sodium poly(styrene sulfonate).

To compute conformational properties, we employ Monte Carlo simulations,
which makes it possible to accelerate the relaxation of longer chains
via reptation moves. Since we focus on the dilute polyelectrolyte
regime, relatively large simulation cells are required, resulting in
considerable ion numbers in systems with added salt. For computational
efficiency, we therefore confine ourselves to a monomer concentration
$C_m=0.008 \sigma^{-3}$, which corresponds to $0.99 {\rm
mol/\ell}$. This is 
higher than typical experimental conditions~\cite{olvera95}, but we have
explicitly verified that our findings are representative for the entire
concentration range $0.001 \sigma^{-3} \leq C_m \leq 0.008 \sigma^{-3}$.
In addition, for all chain lengths studied here ($N=16$, $32$, $64$,
$96$), $C_m$ lies well below the overlap threshold $C_m^*$ as estimated
from the radius of gyration~$R_g$. We adopt $\sigma$, $\sigma^{-3}$, and
$e$ as our units of length, concentration, and charge, respectively.

\begin{figure}
\begin{center}
  \includegraphics[angle=270,width=\figurewidth]{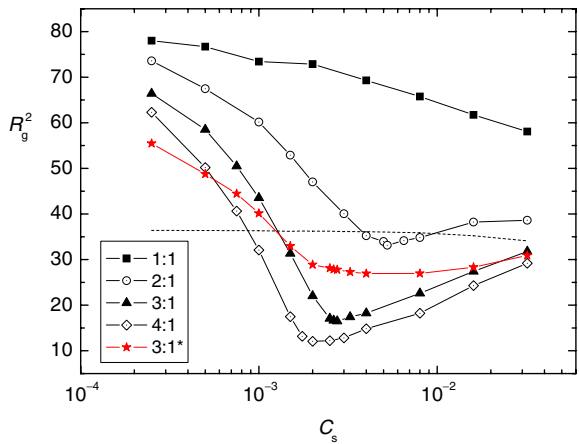}
  \caption{Squared radius of gyration $R_g^2$ of a polyelectrolyte
  of length $N=64$ as a function of salt concentration~$C_s$.
  Curves corresponds to added salt with counterion valency from $1$
  to~$4$. The curve $3$:$1^*$ refers to a system with ions that have a
  twice smaller diameter (see text for details). The dashed line
  corresponds to a reference system without any electrostatic
  interactions. Error bars are smaller than the symbol size.}
  \label{fig:Rg2_Zdependence}
\end{center}
\end{figure}

We first study the conformation of a polyelectrolyte of length $N=64$ in
the presence of ($Z$:$1$)-salt, where $Z$ is varied from 1 to 4. Ions
and monomers have the same size.  Figure~\ref{fig:Rg2_Zdependence} shows
the average squared radius of gyration~$R_g^2$ as a function of salt
concentration. In the absence of salt, the chain adopts an extended
conformation, owing to the electrostatic repulsions between monomers.
Upon addition of monovalent salt, these repulsions are screened and
$\smash{R_g^2}$ gradually decreases.  By contrast, for multivalent salt,
a much stronger decrease in $R_g^2$ is seen, occurring at considerably
lower salt concentrations. This is the conformational collapse that has
previously been observed for polyelectrolytes with multivalent
counterions under salt-free conditions~\cite{winkler98,liu03}.  The
smallest value of $R_g^2$ occurs for $C_s$ near $C_Z$, the
($Z$:$1$)-salt concentration at which the total charge of the $Z$-valent
cations neutralizes the bare polyelectrolyte charge.  Accordingly, this
compact state occurs at a salt concentration that decreases with
increasing valency~$Z$, consistent with the two-state
model~\cite{solis00,solis01}. In addition, higher counterion valency
leads to more compact states. A striking effect occurs once the salt
concentration is increased \emph{beyond} $C_Z$. The chain starts to
swell again, in close analogy with the redissolution observed for
multichain aggregates~\cite{olvera95,pelta96a}. Although the chains that
are collapsed by trivalent and tetravalent counterions reexpand
strongly, they remain more compact than the chain in $2$:$1$ salt for
all salt concentrations investigated.  Comparison with the structure of
the same polyelectrolyte in monovalent salt, which exhibits a slow,
monotonic decrease of $R_g^2$ with increasing $C_s$, emphasizes the
important role of counterion valency.
The salt concentration at which the reexpanded state is reached is
insensitive to monomer concentration, whereas the collapse concentration
$C_Z$ is proportional to~$C_m$. Thus, our relatively high choice of
$C_m$ causes both conformational changes to occur over a narrower range in
salt concentration than in typical experiments.

\begin{figure}[b]
\begin{center}
\includegraphics[angle=270,width=\figurewidth]{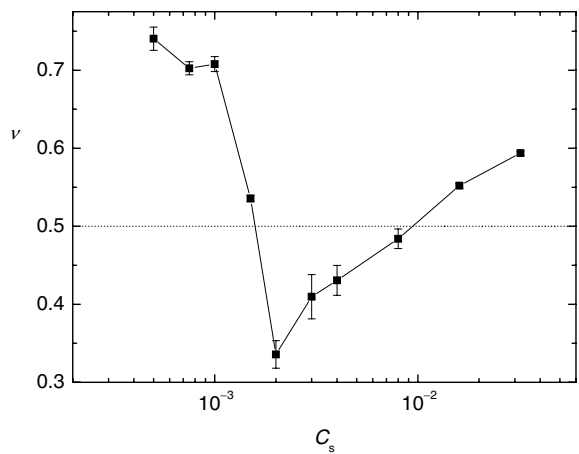}
\caption{Scaling dimension~$\nu$ of the radius of gyration for a
polyelectrolyte in a tetravalent salt solution. At low salt
concentration~$C_s$, $\nu$ exceeds the self-avoiding random-walk
value~$0.588$, owing to the electrostatic repulsions between monomers.
For the concentration~$C_{Z=4}$ corresponding to the minimum in
Fig.~\ref{fig:Rg2_Zdependence}, the exponent reaches the value for a
compact globule. The subsequent reexpansion of the chain at higher salt
concentrations is reflected in an increase in $\nu$.}
\label{fig:nu_Cs}
\end{center}
\end{figure}

We further quantify the swelling of the polyelectrolyte coil by
investigating the variation of $R_g$ with degree of polymerization~$N$,
for tetravalent salt. As anticipated from scaling
theory~\cite{pfeuty77}, the radius of gyration is described by a power
law $N^\nu$ for the range of chain lengths studied here ($16 \leq N \leq
96$). Figure~\ref{fig:nu_Cs} depicts the concentration-dependent
exponent~$\nu$ extracted from least-squares fits at fixed~$C_s$.  At low
concentrations, $\nu$ exceeds the self-avoiding random-walk value
$\nu_{\rm SAW}=0.588$, owing to the electrostatic repulsions between
monomers~\cite{note-exp}.  Near the neutralization
concentration~$C_{Z=4}=0.002$, the chain attains its most compact state
and $\nu=0.34 \pm 0.02$, in good agreement with the poor-solvent value,
$\nu = \smash{\frac{1}{3}}$.  Upon reexpansion, the exponent $\nu$
increases and approximately reaches $\nu_{\rm SAW}$ at the highest
concentrations.

Chain reexpansion has been linked to \emph{charge reversal} (or
\emph{overcharging})~\cite{grosberg02}. Since for spherical colloids
overcharging was found to strongly depend on ion excluded
volume~\cite{messina02}, we proceed to clarify the effect of ion size in
the case of flexible polyelectrolytes. To this end, we repeat our
calculations for $3$:$1$ salt in which all ions have a diameter $\sigma/2$,
i.e., twice smaller than the monomers.  The electrostatic attraction
between a trivalent counterion and a monomer at contact thus increases
by a factor $\frac{4}{3}$ and equals the attraction experienced by
tetravalent ions of diameter~$\sigma$.  As the smaller ions allow the
chain to adopt a more compact structure, it may appear plausible that in
this situation $R_g$ is even smaller than for tetravalent salt and hence
also smaller than $R_g$ for trivalent salt of diameter~$\sigma$.
However, as shown in Fig.~\ref{fig:Rg2_Zdependence} (cf.\ the curve
labeled~$3$:$1^*$), the actual situation is more complicated. The
polyelectrolyte chain is indeed less extended than in tetravalent salt
solution for low salt concentrations and shrinks further if $C_s$ is
increased. But this decrease is weaker than for larger ions, and near
the neutralization condition the chain is much \emph{less compact} than
for tetravalent or even trivalent salt of diameter~$\sigma$. For higher
salt concentrations we observe a gradual expansion of the chain, and
$R_g$ reaches a magnitude that is approximately the same for both
trivalent cases.

\begin{figure}
\begin{center}
  \includegraphics[angle=270,width=\figurewidth]{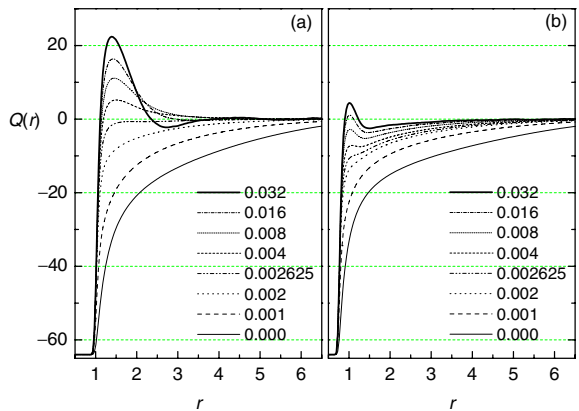}
  \caption{Net charge~$Q(r)$ within a tube of radius~$r$ around a chain
  ($N=64$) at various salt concentrations $C_s$ of trivalent salt.
  (a) For ion diameter~$\sigma$ (case $3$:$1$ in
  Fig.~\ref{fig:Rg2_Zdependence}) an overcharging peak appears at
  short~$r$. (b) For smaller ions (case $3$:$1^*$) the overcharging peak
  is absent.}
  \label{fig:QexVol}
\end{center}
\end{figure}

The mechanism for this remarkable dependence on ion size can be
understood from the charge distribution around the polyelectrolyte. We
define a worm-like tube around the chain, consisting of $N$ spheres of
radius~$r$ centered around the monomers, and monitor the
ensemble-averaged net charge~$Q(r)$ within this tube, as a function of
salt concentration. As shown in Fig.~\ref{fig:QexVol}(a) for trivalent
salt with ion size~$\sigma$, $Q(r)$ is always negative for $0 \leq C_s <
C_{Z=3} = 0.00267$ and only approaches zero at large~$r$, to satisfy
electroneutrality. In this regime, upon addition of salt all trivalent
counterions condense on the chain and replace the monovalent
counterions, as predicted~\cite{solis00}. As $C_s$ approaches $C_{Z=3}$
the magnitude of the net charge $Q(r)$ decreases, until the chain is
completely neutralized by the condensed trivalent counterions and takes
a compact structure. For $C_s > C_{Z=3}$, not all trivalent ions
condense, although the number of condensed trivalent counterions can
exceed the number required to neutralize the chain, in accordance with
Refs.~\cite{nguyen00,grosberg02}. This \emph{overcharging} is reflected
in the appearance of a strong peak at small tube diameter~$r$ in
Fig.~\ref{fig:QexVol}(a). The excess charge attracts coions that form a
second layer, resulting in a rapid decrease of the peak with
increasing~$r$. At the highest salt concentration investigated, $Q(r)$
exhibits an oscillatory tail, indicative of a multi-layered ionic
structure, similar to what has been observed for electrical double
layers~\cite{greberg98}, colloids~\cite{terao01} and rod-like
polyelectrolytes~\cite{deserno02}. Interestingly, for
monovalent salt (not shown) no overcharging peak occurs, consistent with
the absence of important conformational changes when the salt
concentration is varied.

The behavior of~$Q(r)$ drastically changes upon reduction of the ion
size, as illustrated in Fig.~\ref{fig:QexVol}(b) for case $3$:$1^*$ (ion
diameter~$\sigma/2$). Again, the net charge within the tube decreases
with increasing salt concentration, but it is \emph{not} neutralized at
$C_{Z=3}$, consistent with a much larger coil size than for the system
with larger ions (Fig.~\ref{fig:Rg2_Zdependence}). For $C_s > C_{Z=3}$ a
small peak appears in $Q(r)$, but there is hardly any overcharging even
at the highest concentrations. A study of the ion configurations in the
simulations reveals that this remarkable difference arises from the
strong increase in counterion--coion attraction if their size is
reduced. This association, akin to the formation of Bjerrum pairs,
reduces the effective valency of the salt. Consequently, compared to the
$3$:$1$ case, the chain exhibits a less strong conformational collapse
and $Q(r)$ remains more negative. When pairs of trivalent counterions
and monovalent coions condense on the chain, they tend to be oriented
with the trivalent ion toward the monomer, explaining the appearance of
a small peak in~$Q(r)$. Analogous behavior has been observed for simple
colloids~\cite{messina02}, indicating that the decrease in overcharging
upon reduction of ion size is a generic effect.  It also explains why
neither reexpansion nor overcharging of a polyelectrolyte were observed
in earlier simulations of a flexible chain in the presence of divalent
salt~\cite{liu03}. In those simulations the ion sizes were chosen as
half the monomer size~\cite{note-size}.

\begin{figure}
\begin{center}
  \includegraphics[angle=270,width=\figurewidth]{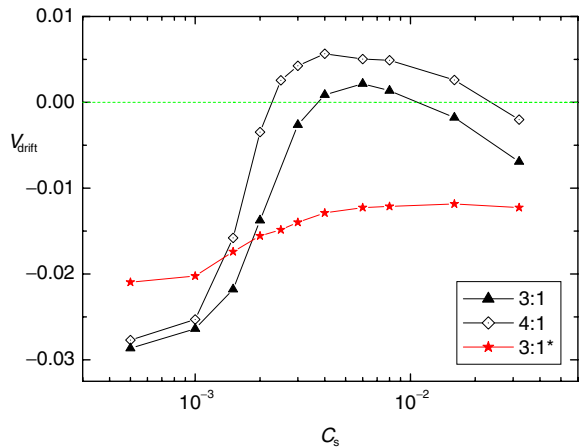}
  \caption{Drift velocity $V_{\rm drift}$ (in units of $\sigma/\tau$) of
  the center of mass of a polyelectrolyte ($N=16$) as a function of salt
  concentration $C_s$,
  in the direction of the electric field. The sign reversal in the
  electrophoretic mobility provides direct evidence for charge reversal
  by multivalent counterions.}
  \label{fig:Vdrift}
\end{center}
\end{figure}

It is tempting to assign an ``effective'' charge to the chain based upon
the net tube charge~$Q(r)$. However, as Fig.~\ref{fig:QexVol} indicates,
such an interpretation is arbitrary if there is no clear prescription
for the proper choice of the tube radius. To circumvent this ambiguity,
we study the electrophoretic mobility as a function of salt valency and
concentration for a polyelectrolyte of $N=16$ monomers via Langevin
dynamics simulations, using the LAMMPS package~\cite{plimpton95}. The
collision frequency in the thermostat equals $\gamma=\tau^{-1}$, where
$\tau=\sigma\sqrt{m/(k_{\rm B}T)}$ is the time unit and $m$ the ion
mass.  We apply a constant electric field of magnitude $E=0.1 k_{\rm
B}T/(e\sigma)$, which, as we carefully verified, does not distort the
conformational properties of the chain or the charge distribution around
it~\cite{note-long}. Due to the absence of hydrodynamic effects this
calculation does not realistically reproduce dynamic properties, but it will
properly reflect the sign of the electrophoretic mobility and can thus
be used to probe the effective charge without imposing an arbitrary
criterion for the condensation of specific ions. Indeed, the
center-of-mass drift velocity of the chain (Fig.~\ref{fig:Vdrift})
reveals a striking dependence on salt concentration. For both trivalent
and tetravalent salt, it exhibits a zero crossing, i.e., the
electrophoretic mobility of the chain \emph{changes sign} upon addition
of salt, as observed experimentally for colloids~\cite{martin-molina03}.
This sign reversal occurs close to the neutralization condition (minimum
in Fig.~\ref{fig:Rg2_Zdependence}) and thus depends on counterion
valency. Furthermore, for trivalent salt with smaller ion size (case
$3$:$1^*$) \emph{no} sign reversal occurs, i.e., no overcharging takes
place. We view these observations as a strong confirmation of the charge
reversal suggested by Fig.~\ref{fig:QexVol} and as evidence for the
relation between charge reversal and reexpansion, but also note that a
less strong collapse and subsequent reexpansion can occur \emph{without}
overcharging.  At the highest salt concentrations, a second sign
reversal occurs, which is possibly related to the formation of a
multi-layered structure of counterions and coions, consistent with the
oscillatory behavior in Fig.~\ref{fig:QexVol}~\cite{note-long}.

In summary, we have demonstrated, for the first time, that dilute
flexible polyelectrolytes in a solution with multivalent salt undergo
not only a conformational collapse, but also a reexpansion.  These
findings support the two-state model for polyelectrolyte
conformations~\cite{solis00,solis01} and confirm the important role of
ion excluded volume~\cite{solis02}. Furthermore, we have presented
unambiguous evidence for overcharging by means of condensed multivalent
counterions at high salt concentrations~\cite{nguyen00,grosberg02}, in
which the reversal of the effective charge coincides with the most
compact chain conformation.

\begin{acknowledgments}
  This material is based upon work supported by the U.S. Department of
  Energy, Division of Materials Sciences under Grant No.\
  DEFG02-91ER45439, through the Frederick Seitz Materials Research
  Laboratory at the University of Illinois at Urbana-Champaign.
\end{acknowledgments}


\begin{thebibliography}{28}
\expandafter\ifx\csname natexlab\endcsname\relax\def\natexlab#1{#1}\fi
\expandafter\ifx\csname bibnamefont\endcsname\relax
  \def\bibnamefont#1{#1}\fi
\expandafter\ifx\csname bibfnamefont\endcsname\relax
  \def\bibfnamefont#1{#1}\fi
\expandafter\ifx\csname citenamefont\endcsname\relax
  \def\citenamefont#1{#1}\fi
\expandafter\ifx\csname url\endcsname\relax
  \def\url#1{\texttt{#1}}\fi
\expandafter\ifx\csname urlprefix\endcsname\relax\def\urlprefix{URL }\fi
\providecommand{\bibinfo}[2]{#2}
\providecommand{\eprint}[2][]{\url{#2}}

\bibitem[{\citenamefont{Olvera de~la Cruz et~al.}(1995)\citenamefont{Olvera
  de~la Cruz, Belloni, Delsanti, Dalbiez, Spalla, and Drifford}}]{olvera95}
\bibinfo{author}{\bibfnamefont{M.}~\bibnamefont{Olvera de~la Cruz}},
  \bibinfo{author}{\bibfnamefont{L.}~\bibnamefont{Belloni}},
  \bibinfo{author}{\bibfnamefont{M.}~\bibnamefont{Delsanti}},
  \bibinfo{author}{\bibfnamefont{J.~P.} \bibnamefont{Dalbiez}},
  \bibinfo{author}{\bibfnamefont{O.}~\bibnamefont{Spalla}}, \bibnamefont{and}
  \bibinfo{author}{\bibfnamefont{M.}~\bibnamefont{Drifford}},
  \bibinfo{journal}{J. Chem. Phys.} \textbf{\bibinfo{volume}{103}},
  \bibinfo{pages}{5781} (\bibinfo{year}{1995}).

\bibitem[{\citenamefont{Pelta et~al.}(1996)\citenamefont{Pelta, Livolant, and
  Sikorav}}]{pelta96a}
\bibinfo{author}{\bibfnamefont{J.}~\bibnamefont{Pelta}},
  \bibinfo{author}{\bibfnamefont{F.}~\bibnamefont{Livolant}}, \bibnamefont{and}
  \bibinfo{author}{\bibfnamefont{J.-L.} \bibnamefont{Sikorav}},
  \bibinfo{journal}{J. Biol. Chem.} \textbf{\bibinfo{volume}{271}},
  \bibinfo{pages}{5656} (\bibinfo{year}{1996}).

\bibitem[{\citenamefont{Raspaud et~al.}(1998)\citenamefont{Raspaud, Olvera
  de~la Cruz, Sikorav, and Livolant}}]{raspaud98}
\bibinfo{author}{\bibfnamefont{E.}~\bibnamefont{Raspaud}},
  \bibinfo{author}{\bibfnamefont{M.}~\bibnamefont{Olvera de~la Cruz}},
  \bibinfo{author}{\bibfnamefont{J.-L.} \bibnamefont{Sikorav}},
  \bibnamefont{and} \bibinfo{author}{\bibfnamefont{F.}~\bibnamefont{Livolant}},
  \bibinfo{journal}{Biophys. J.} \textbf{\bibinfo{volume}{74}},
  \bibinfo{pages}{381} (\bibinfo{year}{1998}).

\bibitem[{\citenamefont{Sabbagh et~al.}(1999)\citenamefont{Sabbagh, Delsanti,
  and Lesieur}}]{sabbagh99}
\bibinfo{author}{\bibfnamefont{I.}~\bibnamefont{Sabbagh}},
  \bibinfo{author}{\bibfnamefont{M.}~\bibnamefont{Delsanti}}, \bibnamefont{and}
  \bibinfo{author}{\bibfnamefont{P.}~\bibnamefont{Lesieur}},
  \bibinfo{journal}{Eur. Phys. J. B} \textbf{\bibinfo{volume}{12}},
  \bibinfo{pages}{253} (\bibinfo{year}{1999}).

\bibitem[{\citenamefont{Sanders et~al.}(2005)\citenamefont{Sanders,
  Gu{\'a}queta, Angelini, Lee, Slimmer, Luijten, and Wong}}]{sanders05}
\bibinfo{author}{\bibfnamefont{L.~K.} \bibnamefont{Sanders}},
  \bibinfo{author}{\bibfnamefont{C.}~\bibnamefont{Gu{\'a}queta}},
  \bibinfo{author}{\bibfnamefont{T.~E.} \bibnamefont{Angelini}},
  \bibinfo{author}{\bibfnamefont{J.-W.} \bibnamefont{Lee}},
  \bibinfo{author}{\bibfnamefont{S.~C.} \bibnamefont{Slimmer}},
  \bibinfo{author}{\bibfnamefont{E.}~\bibnamefont{Luijten}}, \bibnamefont{and}
  \bibinfo{author}{\bibfnamefont{G.~C.~L.} \bibnamefont{Wong}},
  \bibinfo{journal}{Phys. Rev. Lett.} \textbf{\bibinfo{volume}{95}},
  \bibinfo{pages}{108302} (\bibinfo{year}{2005}).

\bibitem[{\citenamefont{Nguyen et~al.}(2000)\citenamefont{Nguyen, Rouzina, and
  Shklovskii}}]{nguyen00}
\bibinfo{author}{\bibfnamefont{T.~T.} \bibnamefont{Nguyen}},
  \bibinfo{author}{\bibfnamefont{I.}~\bibnamefont{Rouzina}}, \bibnamefont{and}
  \bibinfo{author}{\bibfnamefont{B.~I.} \bibnamefont{Shklovskii}},
  \bibinfo{journal}{J. Chem. Phys.} \textbf{\bibinfo{volume}{112}},
  \bibinfo{pages}{2562} (\bibinfo{year}{2000}).

\bibitem[{\citenamefont{Vijayanathan et~al.}(2002)\citenamefont{Vijayanathan,
  Thomas, and Thomas}}]{vijayanathan02}
\bibinfo{author}{\bibfnamefont{V.}~\bibnamefont{Vijayanathan}},
  \bibinfo{author}{\bibfnamefont{T.}~\bibnamefont{Thomas}}, \bibnamefont{and}
  \bibinfo{author}{\bibfnamefont{T.~J.} \bibnamefont{Thomas}},
  \bibinfo{journal}{Biochemistry} \textbf{\bibinfo{volume}{41}},
  \bibinfo{pages}{14085} (\bibinfo{year}{2002}).

\bibitem[{\citenamefont{Murayama et~al.}(2003)\citenamefont{Murayama, Sakamaki,
  and Sano}}]{murayama03}
\bibinfo{author}{\bibfnamefont{Y.}~\bibnamefont{Murayama}},
  \bibinfo{author}{\bibfnamefont{Y.}~\bibnamefont{Sakamaki}}, \bibnamefont{and}
  \bibinfo{author}{\bibfnamefont{M.}~\bibnamefont{Sano}},
  \bibinfo{journal}{Phys. Rev. Lett.} \textbf{\bibinfo{volume}{90}},
  \bibinfo{pages}{018102} (\bibinfo{year}{2003}).

\bibitem[{\citenamefont{Solis and Olvera de~la Cruz}(2000)}]{solis00}
\bibinfo{author}{\bibfnamefont{F.~J.} \bibnamefont{Solis}} \bibnamefont{and}
  \bibinfo{author}{\bibfnamefont{M.}~\bibnamefont{Olvera de~la Cruz}},
  \bibinfo{journal}{J. Chem. Phys.} \textbf{\bibinfo{volume}{112}},
  \bibinfo{pages}{2030} (\bibinfo{year}{2000}).

\bibitem[{\citenamefont{Solis and Olvera de~la Cruz}(2001)}]{solis01}
\bibinfo{author}{\bibfnamefont{F.~J.} \bibnamefont{Solis}} \bibnamefont{and}
  \bibinfo{author}{\bibfnamefont{M.}~\bibnamefont{Olvera de~la Cruz}},
  \bibinfo{journal}{Eur. Phys. J. E} \textbf{\bibinfo{volume}{4}},
  \bibinfo{pages}{143} (\bibinfo{year}{2001}).

\bibitem[{\citenamefont{Solis}(2002)}]{solis02}
\bibinfo{author}{\bibfnamefont{F.~J.} \bibnamefont{Solis}},
  \bibinfo{journal}{J. Chem. Phys.} \textbf{\bibinfo{volume}{117}},
  \bibinfo{pages}{9009} (\bibinfo{year}{2002}).

\bibitem[{\citenamefont{Grosberg et~al.}(2002)\citenamefont{Grosberg, Nguyen,
  and Shklovskii}}]{grosberg02}
\bibinfo{author}{\bibfnamefont{A.~Y.} \bibnamefont{Grosberg}},
  \bibinfo{author}{\bibfnamefont{T.~T.} \bibnamefont{Nguyen}},
  \bibnamefont{and} \bibinfo{author}{\bibfnamefont{B.~I.}
  \bibnamefont{Shklovskii}}, \bibinfo{journal}{Rev. Mod. Phys.}
  \textbf{\bibinfo{volume}{74}}, \bibinfo{pages}{329} (\bibinfo{year}{2002}).

\bibitem[{\citenamefont{Stevens and Plimpton}(1998)}]{stevens98}
\bibinfo{author}{\bibfnamefont{M.~J.} \bibnamefont{Stevens}} \bibnamefont{and}
  \bibinfo{author}{\bibfnamefont{S.~J.} \bibnamefont{Plimpton}},
  \bibinfo{journal}{Eur. Phys. J. B} \textbf{\bibinfo{volume}{2}},
  \bibinfo{pages}{341} (\bibinfo{year}{1998}).

\bibitem[{\citenamefont{Liu et~al.}(2003)\citenamefont{Liu, Ghosh, and
  Muthukumar}}]{liu03}
\bibinfo{author}{\bibfnamefont{S.}~\bibnamefont{Liu}},
  \bibinfo{author}{\bibfnamefont{K.}~\bibnamefont{Ghosh}}, \bibnamefont{and}
  \bibinfo{author}{\bibfnamefont{M.}~\bibnamefont{Muthukumar}},
  \bibinfo{journal}{J. Chem. Phys.} \textbf{\bibinfo{volume}{119}},
  \bibinfo{pages}{1813} (\bibinfo{year}{2003}).

\bibitem[{\citenamefont{Stevens and Kremer}(1995)}]{stevens95}
\bibinfo{author}{\bibfnamefont{M.~J.} \bibnamefont{Stevens}} \bibnamefont{and}
  \bibinfo{author}{\bibfnamefont{K.}~\bibnamefont{Kremer}},
  \bibinfo{journal}{J. Chem. Phys.} \textbf{\bibinfo{volume}{103}},
  \bibinfo{pages}{1669} (\bibinfo{year}{1995}).

\bibitem[{\citenamefont{Winkler et~al.}(1998)\citenamefont{Winkler, Gold, and
  Reineker}}]{winkler98}
\bibinfo{author}{\bibfnamefont{R.~G.} \bibnamefont{Winkler}},
  \bibinfo{author}{\bibfnamefont{M.}~\bibnamefont{Gold}}, \bibnamefont{and}
  \bibinfo{author}{\bibfnamefont{P.}~\bibnamefont{Reineker}},
  \bibinfo{journal}{Phys. Rev. Lett.} \textbf{\bibinfo{volume}{80}},
  \bibinfo{pages}{3731} (\bibinfo{year}{1998}).

\bibitem[{\citenamefont{Stevens}(2001)}]{stevens01}
\bibinfo{author}{\bibfnamefont{M.~J.} \bibnamefont{Stevens}},
  \bibinfo{journal}{Biophys. J.} \textbf{\bibinfo{volume}{80}},
  \bibinfo{pages}{130} (\bibinfo{year}{2001}).

\bibitem[{\citenamefont{Limbach and Holm}(2003)}]{limbach03}
\bibinfo{author}{\bibfnamefont{H.~J.} \bibnamefont{Limbach}} \bibnamefont{and}
  \bibinfo{author}{\bibfnamefont{C.}~\bibnamefont{Holm}}, \bibinfo{journal}{J.
  Phys. Chem. B} \textbf{\bibinfo{volume}{107}}, \bibinfo{pages}{8041}
  (\bibinfo{year}{2003}).

\bibitem[{\citenamefont{Pfeuty et~al.}(1977)\citenamefont{Pfeuty, Velasco, and
  de~Gennes}}]{pfeuty77}
\bibinfo{author}{\bibfnamefont{P.}~\bibnamefont{Pfeuty}},
  \bibinfo{author}{\bibfnamefont{R.~M.} \bibnamefont{Velasco}},
  \bibnamefont{and} \bibinfo{author}{\bibfnamefont{P.~G.}
  \bibnamefont{de~Gennes}}, \bibinfo{journal}{J. Phys. (Paris) Lett.}
  \textbf{\bibinfo{volume}{38}}, \bibinfo{pages}{5} (\bibinfo{year}{1977}).

\bibitem[{not({\natexlab{a}})}]{note-exp}
\bibinfo{note}{The effective value of~$\nu$ in the absence of salt sensitively
  depends on the monomer concentration.}

\bibitem[{\citenamefont{Messina et~al.}(2002)\citenamefont{Messina,
  Gonz\'alez-Tovar, Lozada-Cassou, and Holm}}]{messina02}
\bibinfo{author}{\bibfnamefont{R.}~\bibnamefont{Messina}},
  \bibinfo{author}{\bibfnamefont{E.}~\bibnamefont{Gonz\'alez-Tovar}},
  \bibinfo{author}{\bibfnamefont{M.}~\bibnamefont{Lozada-Cassou}},
  \bibnamefont{and} \bibinfo{author}{\bibfnamefont{C.}~\bibnamefont{Holm}},
  \bibinfo{journal}{Europhys. Lett.} \textbf{\bibinfo{volume}{60}},
  \bibinfo{pages}{383} (\bibinfo{year}{2002}).

\bibitem[{\citenamefont{Greberg and Kjellander}(1998)}]{greberg98}
\bibinfo{author}{\bibfnamefont{H.}~\bibnamefont{Greberg}} \bibnamefont{and}
  \bibinfo{author}{\bibfnamefont{R.}~\bibnamefont{Kjellander}},
  \bibinfo{journal}{J. Chem. Phys.} \textbf{\bibinfo{volume}{108}},
  \bibinfo{pages}{2940} (\bibinfo{year}{1998}).

\bibitem[{\citenamefont{Terao and Nakayama}(2001)}]{terao01}
\bibinfo{author}{\bibfnamefont{T.}~\bibnamefont{Terao}} \bibnamefont{and}
  \bibinfo{author}{\bibfnamefont{T.}~\bibnamefont{Nakayama}},
  \bibinfo{journal}{Phys. Rev. E} \textbf{\bibinfo{volume}{63}},
  \bibinfo{pages}{041401} (\bibinfo{year}{2001}).

\bibitem[{\citenamefont{Deserno and Holm}(2002)}]{deserno02}
\bibinfo{author}{\bibfnamefont{M.}~\bibnamefont{Deserno}} \bibnamefont{and}
  \bibinfo{author}{\bibfnamefont{C.}~\bibnamefont{Holm}},
  \bibinfo{journal}{Mol. Phys.} \textbf{\bibinfo{volume}{100}},
  \bibinfo{pages}{2941} (\bibinfo{year}{2002}).

\bibitem[{not({\natexlab{b}})}]{note-size}
\bibinfo{note}{For the conditions examined here, this size would be
  unrealistically small, if one takes into account hydration of the ions.}

\bibitem[{\citenamefont{Plimpton}(1995)}]{plimpton95}
\bibinfo{author}{\bibfnamefont{S.~J.} \bibnamefont{Plimpton}},
  \bibinfo{journal}{J. Comp. Phys.} \textbf{\bibinfo{volume}{117}},
  \bibinfo{pages}{1} (\bibinfo{year}{1995}).

\bibitem[{not({\natexlab{c}})}]{note-long}
\bibinfo{note}{P.-Y. Hsiao and E. Luijten, to be published.}

\bibitem[{\citenamefont{Martin-Molina et~al.}(2003)\citenamefont{Martin-Molina,
  Quesada-P\'erez, Galisteo-Gonz\'alez, and
  Hidalgo-\'Alvarez}}]{martin-molina03}
\bibinfo{author}{\bibfnamefont{A.}~\bibnamefont{Martin-Molina}},
  \bibinfo{author}{\bibfnamefont{M.}~\bibnamefont{Quesada-P\'erez}},
  \bibinfo{author}{\bibfnamefont{F.}~\bibnamefont{Galisteo-Gonz\'alez}},
  \bibnamefont{and}
  \bibinfo{author}{\bibfnamefont{R.}~\bibnamefont{Hidalgo-\'Alvarez}},
  \bibinfo{journal}{J. Chem. Phys.} \textbf{\bibinfo{volume}{118}},
  \bibinfo{pages}{4183} (\bibinfo{year}{2003}).

\end{thebibliography}

\end{document}